\documentclass[usenatbib,letters]{mnras}

\usepackage[utf8]{inputenc}

\usepackage{newtxtext,newtxmath}
\usepackage[T1]{fontenc}
\usepackage{bm}

\renewcommand{\vec}{\bm}
\newcommand{\mat}{\mathbfss}
\newcommand{\I}{\mathop{}\!\mathrm{i}\!\mathop{}}
\newcommand{\Norm}{\mathop{\mathcal{N}}}
\newcommand{\E}{\operatorname{E}}
\newcommand{\Var}{\operatorname{Var}}
\newcommand{\hVar}{\operatorname{\hat Var}}

\newcommand{\hCov}{\operatorname{\hat Cov}}
\newcommand{\erfc}{\operatorname{erfc}}
\newcommand{\tp}{\mathrm{T}}

\newcommand{\matSigma}{\bmath \Sigma}

\title[An unbiased estimator for the ellipticity]{An unbiased estimator for the ellipticity from image moments}

\author[N. Tessore]{%
Nicolas Tessore%
\thanks{Email: \url{nicolas.tessore@manchester.ac.uk}}
\\
Jodrell Bank Centre for Astrophysics, University of Manchester,
Alan Turing Building, Oxford Road, Manchester, M13 9PL, UK
}

\date{Accepted 2017 June 14. Received 2017 May 31; in original form 2017 May 07}

\pubyear{2017}

\begin{document}
\label{firstpage}
\pagerange{\pageref{firstpage}--\pageref{lastpage}}
\maketitle

\begin{abstract}
An unbiased estimator for the ellipticity of an object in a noisy image is given in terms of the image moments.
Three assumptions are made: i) the pixel noise is normally distributed, although with arbitrary covariance matrix, ii) the image moments are taken about a fixed centre, and iii) the point-spread function is known.
The relevant combinations of image moments are then jointly normal and their covariance matrix can be computed.
A particular estimator for the ratio of the means of jointly normal variates is constructed and used to provide the unbiased estimator for the ellipticity.
Furthermore, an unbiased estimate of the covariance of the new estimator is also given.
\end{abstract}


\section{Introduction}

A number of applications in astronomy require the measurement of shapes of objects from observed images.
A commonly used shape descriptor is the ellipticity~$\chi$, which is defined in terms of the central image moments~$m_{pq}$ as the complex number
\begin{equation}\label{eq:chi-def}
	\chi
	= \chi_1 + \I \chi_2
	= \frac{m_{20} - m_{02} + 2 \I m_{11}}{m_{20} + m_{02}} \;.
\end{equation}
For example, in weak gravitational lensing, the gravitational field distorts the observed shapes of background galaxies, and this shear can be detected in ellipticity measurements.
This is possible because the observed ellipticity of a source affected by gravitational lensing is \citep[see e.g.\@][]{2001PhR...340..291B}
\begin{equation}
	\chi = \frac{\chi_\text{s} + 2 g + g^2 \chi_\text{s}^*}{1 + |g|^2 + 2 \Re[g \chi_\text{s}^*]} \;,
\end{equation}
where $\chi_\text{s}$ is the intrinsic ellipticity of the source, and $g = g_1 + \I g_2$ is the so-called \emph{reduced shear} of the gravitational lens.
In the limit of weak lensing, this can be approximated to linear order as
\begin{equation}\label{eq:wl}
	\chi \approx \chi_\text{s} + 2 g \;.
\end{equation}
Averaging over an ensemble of randomly-oriented background sources, i.e.\@ $\langle \chi_\text{s} \rangle = 0$, the weak lensing equation~\eqref{eq:wl} yields
\begin{equation}\label{eq:1pt}
	\langle \chi \rangle = 2 \langle g \rangle \;.
\end{equation}
The observed ellipticities are thus a direct estimator for the shear~$g$ from gravitational lensing.
Similar ideas are used in Cosmology \citep[for a recent review, see][]{2015RPPh...78h6901K}.
Here, \emph{cosmic shear} from the large-scale structure of the universe imprints a specific signature onto the ellipticity two-point correlation functions,
\begin{equation}\label{eq:2pt}
	\xi_{ij}(r) = \left\langle \chi_i(\vec x) \, \chi_j(\vec y) \right\rangle_{\vert \vec x - \vec y \vert = r} \;,
\end{equation}
where the average is taken over pairs of sources with the given separation~$r$ on the sky.
Note that both the one-point function~\eqref{eq:1pt} and the two-point function~\eqref{eq:2pt} depend on the mean ellipticity over a potentially large sample of sources.

In practice, an estimator~$\hat \chi$ is used to measure the ellipticities of observed sources.
In order to not introduce systematic errors into applications such as the above, the ellipticity estimator~$\hat \chi$ must be \emph{unbiased}, i.e.\@ $\E[\hat \chi] = \chi$.
One of the biggest problems for estimators that directly work on the data is the \emph{noise bias} \citep{2012MNRAS.425.1951R} arising from pixel noise in the observations.
For example, the standard approach to moment-based shape measurement is to obtain estimates~$\hat m_{pq}$ of the second-order moments from the (noisy) data and use~\eqref{eq:chi-def} directly as an ellipticity estimate,
\begin{equation}\label{eq:hat-e}
	\hat e = \frac{\hat m_{20} - \hat m_{02} + 2 \I \hat m_{11}}{\hat m_{20} + \hat m_{02}} \;.
\end{equation}
The statistical properties of this estimator have been studied by \citet{2012MNRAS.424.2757M} and \citet{2014MNRAS.439.1909V}, who assumed that the image moments are jointly normal with some given variance and correlation.
The estimator~\eqref{eq:hat-e} then follows the distribution of \citet{Marsaglia:1965wx,Marsaglia:2006be} for the ratio of jointly normal variates.
None of the moments of this distribution exist, and even for a finite sample, small values in the denominator can quickly cause significant biases and large variances.
The estimator~\eqref{eq:hat-e} is thus generally poorly behaved, unless the signal-to-noise ratio of the data is very high.

Here, a new unbiased estimator for the ellipticity~$\chi$ from the second-order image moments is proposed.
First, it is shown that for normally-distributed pixel noise with known covariance matrix and a fixed centre, the relevant combinations of image moments are indeed jointly normal and that their covariance matrix can easily be computed.
In the appendix, an unbiased estimator for the ratio of the means of jointly normal variates is constructed, which can subsequently be applied to the image moments.
This produces the ellipticity estimate, as well as unbiased estimates of its variance and covariance.

\section{An unbiased estimator for the ellipticity}
\label{sec:ellipticity}

It is assumed that the data is a random vector $\vec d = (d_1, d_2, \dots)$ of pixels following a multivariate normal distribution centred on the unknown true signal~$\vec \mu$,
\begin{equation}
	\vec d \sim \Norm(\vec \mu, \matSigma) \;,
\end{equation}
where~$\matSigma$ is the covariance matrix for the noise, which is assumed to be known but not restricted to a particular diagonal shape.
The observed signal usually involves a point-spread function (PSF), and it is further assumed that this effect can be approximated as a linear convolution of the true signal and the discretised PSF.
In this case, definition~\eqref{eq:chi-def} can be extended to obtain the true ellipticity of the object before convolution,
\begin{equation}\label{eq:chi-psf}
	\chi
	= \frac{m_{20} - m_{02} - m_{00} \, (\pi_{20} - \pi_{02}) + 2 \I m_{11} - 2 \I m_{00} \, \pi_{11}}{m_{20} + m_{02} - m_{00} \, (\pi_{20} + \pi_{02})} \;,
\end{equation}
from the central moments~$\pi_{pq}$ of the (normalised) PSF.
Fixing a centre~$(\bar x, \bar y)$, the relevant combinations of moments to compute the ellipticity from data~$\vec d$ are thus
\begin{equation}\label{eq:uvs}
\begin{alignedat}{2}
	u &= \sum_i \alpha_i \, d_i \;, \quad &
	\alpha_i &= w_i \, \big((x_i - \bar x)^2 - (y_i - \bar y)^2 - \pi_{20} + \pi_{02}\big) \;, \\
	v &= \sum_i \beta_i \, d_i \;, \quad &
	\beta_i &= w_i \, \big(2 \, (x_i - \bar x) \, (y_i - \bar y) - 2 \, \pi_{11}\big) \;, \\
	s &= \sum_i \gamma_i \, d_i \;, \quad &
	\gamma_i &= w_i \, \big((x_i - \bar x)^2 + (y_i - \bar y)^2 - \pi_{20} - \pi_{02}\big) \;,
\end{alignedat}
\end{equation}
where~$w_i$ is the window function of the observation.
To obtain the true ellipticity estimate of the signal, and for the PSF correction~\eqref{eq:chi-psf} to remain valid, the window function must be unity over the support of the signal.\footnote{%
This is in contrast to the \emph{weight} functions that are sometimes used in moment-based methods to reduce the influence of noise far from the centre.}

Due to the linearity in the pixel values $d_i$, the vector $(u,v,s)$ can be written in matrix form,
\begin{equation}
	(u,v,s) = \mat M \, \vec d \;,
\end{equation}
where the three rows $\alpha_i$, $\beta_i$, $\gamma_i$ of matrix~$\mat M$ are defined by~\eqref{eq:uvs}.
The random vector $(u,v,s)$ is hence normally distributed,
\begin{equation}
	 (u,v,s) \sim \Norm(\vec \mu_{uvs}, \matSigma_{uvs}) \;,
\end{equation}
with unknown mean $\vec \mu_{uvs} = (\mu_u, \mu_v, \mu_s) = \mat M \, \vec \mu$ and known $3 \times 3$ covariance matrix $\matSigma_{uvs} = \mat M \, \matSigma \, \mat M^\tp$ with entries
\begin{equation}\label{eq:Sigma_uvs}
\begin{alignedat}{2}
	 \Sigma_{uu} &= \sum_{i,j} \alpha_i \, \alpha_j \, \Sigma_{ij} \;, \quad &
	 \Sigma_{uv} &= \Sigma_{vu} = \sum_{i,j} \alpha_i \, \beta_j \, \Sigma_{ij} \;, \\
	 \Sigma_{vv} &= \sum_{i,j} \beta_i \, \beta_j \, \Sigma_{ij} \;, \quad &
	 \Sigma_{us} &= \Sigma_{su} = \sum_{i,j} \alpha_i \, \gamma_j \, \Sigma_{ij} \;, \\
	 \Sigma_{ss} &= \sum_{i,j} \gamma_i \, \gamma_j \, \Sigma_{ij} \;, \quad &
	 \Sigma_{vs} &= \Sigma_{sv} = \sum_{i,j} \beta_i \, \gamma_j \, \Sigma_{ij} \;,
\end{alignedat}
\end{equation}
where $\Sigma_{ij}$ are the entries of the covariance matrix~$\matSigma$ of the pixel noise.
Hence the covariance matrix~$\matSigma_{uvs}$ of the moments can be computed if the pixel noise statistics are known.

The true ellipticity~\eqref{eq:chi-psf} of the signal can be written in terms of the mean values of the variates $u$, $v$ and $s$ defined in~\eqref{eq:uvs} as
\begin{equation}\label{eq:chi}
	\chi
	= \chi_1 + \I \chi_2
	= \frac{\mu_u + \I \mu_v}{\mu_s} \;.
\end{equation}
The problem is to find an unbiased estimate of $\chi_1$ and $\chi_2$ from the observed values of $u$, $v$ and $s$.
In appendix~\ref{sec:ratio-estimator}, an unbiased estimator is given for the ratio of the means of two jointly normal random variables.
It can be applied directly to the ellipticity~\eqref{eq:chi}.
First, two new variates~$p$ and~$q$ are introduced,
\begin{equation}\label{eq:pq}
\begin{alignedat}{2}
	p = u - a \, s \;, \quad &
	a = \Sigma_{us}/\Sigma_{ss} \;, \\
	q = v - b \, s \;, \quad &
	b = \Sigma_{vs}/\Sigma_{ss} \;,
\end{alignedat}
\end{equation}
where~$a$, $b$ are constants.
This definition corresponds to~\eqref{eq:w} in the univariate case, and therefore both $(p,s)$ and $(q,s)$ are pairs of independent normal variates.
The desired estimator for the ellipticity is then
\begin{equation}\label{eq:hat-chi}
	\hat \chi_1 = a + p \, \hat g(s) \;, \quad
	\hat \chi_2 = b + q \, \hat g(s) \;.
\end{equation}
Because the mean $\mu_s$ is always positive for a realistic signal, the function~$\hat g(s)$ is given by~\eqref{eq:hat-g}.
From the expectation
\begin{equation}
\begin{alignedat}{4}
	\E[\hat \chi_1]
	&= a + \E[p] \E[\hat g(s)]
	&&= a + (\mu_u - a \, \mu_s)/\mu_s
	&&= \chi_1 \;, \\
	\E[\hat \chi_2]
	&= b + \E[q] \E[\hat g(s)]
	&&= b + (\mu_v - b \, \mu_s)/\mu_s
	&&= \chi_2 \;,
\end{alignedat}
\end{equation}
it follows that $\hat \chi = \hat \chi_1 + \I \hat \chi_2$ is indeed an unbiased estimator for the true ellipticity~$\chi$ of the signal.

In addition, an unbiased estimate of the the variance of~$\hat \chi_1$ and~$\hat \chi_2$ is provided by~\eqref{eq:var-r},
\begin{equation}\label{eq:var-chi}
\begin{aligned}
	\hVar[\hat \chi_1]
	&= p^2 \, \big(\hat g(s)\big)^2
	  - \big((p^2 - \Sigma_{uu})/\Sigma_{ss} + a^2\big) \, 
	    \big(1 - s \, \hat g(s)\big) \;, \\
	\hVar[\hat \chi_2]
	&= q^2 \, \big(\hat g(s)\big)^2
	  - \big((q^2 - \Sigma_{vv})/\Sigma_{ss} + b^2\big) \,
	    \big(1 - s \, \hat g(s)\big) \;.
\end{aligned}	
\end{equation}
Similarly, there is an unbiased estimator for the covariance,
\begin{equation}\label{eq:cov-chi}
	\hCov[\hat \chi_1, \hat \chi_2]
	= pq \, \big(\hat g(s)\big)^2
	  - \big((pq - \Sigma_{uv})/\Sigma_{ss} + ab\big) \, \big(1 - s \, \hat g(s)\big) \;.
\end{equation}
It follows that the individual estimates of the ellipticity components are in general not independent.
However, for realistic pixel noise, window functions and PSFs, the correlations between $u$, $v$ and $s$, and hence $\hat \chi_1$ and $\hat \chi_2$, can become very small.

\begin{table*}%
\caption{Monte Carlo results for the unbiased ellipticity estimator}%
\label{tab:1}%
\centering%
\begin{tabular}{r r c c r r c c c c}%
\hline\hline
\multicolumn{1}{c}{$\chi_1$} &
\multicolumn{1}{c}{$\chi_2$} &
PSF &
counts &
\multicolumn{1}{c}{$\big\langle\hat \chi_1\big\rangle$} &
\multicolumn{1}{c}{$\big\langle\hat \chi_2\big\rangle$} &
$\Var[\hat \chi_1]^{1/2}$ &
$\Var[\hat \chi_2]^{1/2}$ &
$\big\langle{\hVar[\hat \chi_1]}\big\rangle^{1/2}$ &
$\big\langle{\hVar[\hat \chi_2]}\big\rangle^{1/2}$ \\
\hline
$0.1107$ & $0.0000$ & no & 500 & $0.1113\,(08)$ & $0.0006\,(11)$ & $0.762$ & $1.112$ & $0.762$ & $1.111$ \\
&&& 1000 & $0.1105\,(02)$ & $-0.0002\,(02)$ & $0.216$ & $0.214$ & $0.216$ & $0.214$ \\
&&& 2000 & $0.1108\,(01)$ & $0.0000\,(01)$ & $0.104$ & $0.103$ & $0.104$ & $0.103$ \\
&& yes & 500 & $0.1115\,(07)$ & $0.0006\,(06)$ & $0.717$ & $0.598$ & $0.717$ & $0.599$ \\
&&& 1000 & $0.1111\,(02)$ & $-0.0001\,(02)$ & $0.216$ & $0.214$ & $0.215$ & $0.214$ \\
&&& 2000 & $0.1108\,(01)$ & $0.0000\,(01)$ & $0.104$ & $0.103$ & $0.104$ & $0.103$ \\
$0.1820$ & $-0.1842$ & no & 500 & $0.1820\,(07)$ & $-0.1840\,(09)$ & $0.704$ & $0.853$ & $0.704$ & $0.853$ \\
&&& 1000 & $0.1817\,(02)$ & $-0.1842\,(02)$ & $0.225$ & $0.226$ & $0.225$ & $0.226$ \\
&&& 2000 & $0.1822\,(01)$ & $-0.1842\,(01)$ & $0.108$ & $0.108$ & $0.108$ & $0.108$ \\
&& yes & 500 & $0.1802\,(07)$ & $-0.1840\,(06)$ & $0.658$ & $0.616$ & $0.658$ & $0.616$ \\
&&& 1000 & $0.1819\,(02)$ & $-0.1845\,(02)$ & $0.223$ & $0.224$ & $0.224$ & $0.225$ \\
&&& 2000 & $0.1819\,(01)$ & $-0.1841\,(01)$ & $0.108$ & $0.108$ & $0.108$ & $0.108$ \\
$0.0000$ & $0.5492$ & no & 500 & $-0.0005\,(08)$ & $0.5489\,(15)$ & $0.793$ & $1.451$ & $0.794$ & $1.450$ \\
&&& 1000 & $-0.0001\,(02)$ & $0.5494\,(03)$ & $0.248$ & $0.323$ & $0.248$ & $0.322$ \\
&&& 2000 & $-0.0002\,(01)$ & $0.5493\,(01)$ & $0.117$ & $0.149$ & $0.117$ & $0.149$ \\
&& yes & 500 & $0.0001\,(07)$ & $0.5473\,(10)$ & $0.720$ & $0.957$ & $0.720$ & $0.960$ \\
&&& 1000 & $0.0002\,(02)$ & $0.5488\,(03)$ & $0.244$ & $0.315$ & $0.244$ & $0.316$ \\
&&& 2000 & $0.0001\,(01)$ & $0.5491\,(01)$ & $0.116$ & $0.147$ & $0.116$ & $0.147$ \\
\hline
\end{tabular}%
\end{table*}

To demonstrate that the proposed estimator is in fact unbiased under the given assumptions of i) normal pixel noise with known covariance, ii) a fixed centre for the image moments, and iii) the discrete convolution with a known PSF, a Monte Carlo simulation was performed using mock observations of an astronomical object.
The images are postage stamps of $49 \times 49$ pixels, containing a centred source that is truncated near the image borders.
A circular aperture is used as window function.
The source is elliptical and follows the light profile of \citet{1948AnAp...11..247D}.
The ellipse containing half the total light has a 10~pixel semi-major axis and ellipticity as specified.
Where indicated, the signal is convolved with a Gaussian PSF with 5~pixel FWHM.
The pixel noise is uncorrelated with unit variance.
The normalisation $N$ of the object (i.e.\@ the total number of counts) varies to show the effect of the signal-to-noise ratio on the variance of the estimator.\footnote{%
To compare the results to a given data set, it is then possible to scale the data so that the noise has unit variance, and compare the number of counts.
For example, the control-ground-constant data set of the GREAT3 challenge \citep{2014ApJS..212....5M} has mostly $N = 500$--$1000$.
}
The signal ellipticity~$\chi$ is computed from the image before the PSF is applied and noise is added.
The mean of the estimator~$\hat \chi$ is computed from $10^6$ realisations of noise for the same signal.
Also computed are the square root of the sample variance~$\Var[\hat \chi]$ and the mean of the estimated variance $\hVar[\hat \chi]$, respectively.
The results shown in Table~\ref{tab:1} indicate that the estimator performs as expected.

\section{Conclusion \& discussion}
\label{sec:conclusion}

The unbiased estimator~\eqref{eq:hat-chi} provides a new method for ellipticity measurement from noisy images.
Its simple and analytic form allows quick implementation and fast evaluation, and statistics for the results can be obtained directly with unbiased estimates of the variance~\eqref{eq:var-chi} and covariance~\eqref{eq:cov-chi}.

Using an unbiased estimator for the ellipticity of the signal~$\vec \mu$ eliminates the influence of noise from a subsequent analysis of the results (the so-called ``noise bias'' in weak lensing, \citealp{2012MNRAS.425.1951R}).
However, depending on the application, other kinds of biases might exist even for a noise-free image.
For example, due to the discretisation of the image, the signal ellipticity can differ from the intrinsic ellipticity of the observed object.
This ``pixellation bias'' \citep{2016arXiv160907937S} remains an issue in applications such as weak lensing, where the relevant effects must be measured from the intrinsic ellipticity of the objects.

Furthermore, a fixed centre $(\bar x, \bar y)$ for the moments has been assumed throughout.
For a correct ellipticity estimate, this must be the centroid of the signal, which is usually estimated from the data itself.
Centroid errors \citep{2012MNRAS.424.2757M} might therefore ultimately bias the ellipticity estimator or increase its variance, although this currently does not seem to be a significant effect.

In practice, additional biases might arise when the assumed requirements for the window function and PSF are not fulfilled by the data.
The estimator should therefore always be carefully tested for the application at hand.

Lastly, the ellipticity estimate might be improved by suitable filtering of the observed image.
A linear filter with matrix~$\mat A$ can be applied to the image before estimating the ellipticity, since the transformed pixels remain multivariate normal with mean~$\vec \mu' = \mat A \, \vec \mu$ and covariance matrix~$\matSigma' = \mat A \, \matSigma \, \mat A^\tp$.
Examples of viable filters are nearest-neighbour or bilinear interpolation, as well as convolution.
A combination of these filters could be used to perform PSF deconvolution on the observed image, as an alternative to the algebraic PSF correction~\eqref{eq:chi-psf}.

\section*{Acknowledgements}
NT would like to thank S.~Bridle for encouragement and many conversations about shape measurement.
The author acknowledges support from the European Research Council in the form of a Consolidator Grant with number 681431.

\bibliographystyle{mnras}
\bibliography{mnras}

\begin{thebibliography}{}
\makeatletter
\relax
\def\mn@urlcharsother{\let\do\@makeother \do\$\do\&\do\#\do\^\do\_\do\%\do\~}
\def\mn@doi{\begingroup\mn@urlcharsother \@ifnextchar [ {\mn@doi@}
  {\mn@doi@[]}}
\def\mn@doi@[#1]#2{\def\@tempa{#1}\ifx\@tempa\@empty \href
  {http://dx.doi.org/#2} {doi:#2}\else \href {http://dx.doi.org/#2} {#1}\fi
  \endgroup}
\def\mn@eprint#1#2{\mn@eprint@#1:#2::\@nil}
\def\mn@eprint@arXiv#1{\href {http://arxiv.org/abs/#1} {{\tt arXiv:#1}}}
\def\mn@eprint@dblp#1{\href {http://dblp.uni-trier.de/rec/bibtex/#1.xml}
  {dblp:#1}}
\def\mn@eprint@#1:#2:#3:#4\@nil{\def\@tempa {#1}\def\@tempb {#2}\def\@tempc
  {#3}\ifx \@tempc \@empty \let \@tempc \@tempb \let \@tempb \@tempa \fi \ifx
  \@tempb \@empty \def\@tempb {arXiv}\fi \@ifundefined
  {mn@eprint@\@tempb}{\@tempb:\@tempc}{\expandafter \expandafter \csname
  mn@eprint@\@tempb\endcsname \expandafter{\@tempc}}}

\bibitem[\protect\citeauthoryear{{Bartelmann} \& {Schneider}}{{Bartelmann} \&
  {Schneider}}{2001}]{2001PhR...340..291B}
{Bartelmann} M.,  {Schneider} P.,  2001, \mn@doi [\physrep]
  {10.1016/S0370-1573(00)00082-X}, \href
  {http://adsabs.harvard.edu/abs/2001PhR...340..291B} {340, 291}

\bibitem[\protect\citeauthoryear{{Kilbinger}}{{Kilbinger}}{2015}]{2015RPPh...78h6901K}
{Kilbinger} M.,  2015, \mn@doi [Rep. Prog. Phys.]
  {10.1088/0034-4885/78/8/086901}, \href
  {http://adsabs.harvard.edu/abs/2015RPPh...78h6901K} {78, 086901}

\bibitem[\protect\citeauthoryear{{Mandelbaum} et~al.,}{{Mandelbaum}
  et~al.}{2014}]{2014ApJS..212....5M}
{Mandelbaum} R.,  et~al., 2014, \mn@doi [\apjs] {10.1088/0067-0049/212/1/5},
  \href {http://adsabs.harvard.edu/abs/2014ApJS..212....5M} {212, 5}

\bibitem[\protect\citeauthoryear{{Marsaglia}}{{Marsaglia}}{1965}]{Marsaglia:1965wx}
{Marsaglia} G.,  1965, J. Amer. Statist. Assoc., 60, 193

\bibitem[\protect\citeauthoryear{{Marsaglia}}{{Marsaglia}}{2006}]{Marsaglia:2006be}
{Marsaglia} G.,  2006, \mn@doi [J. Stat. Softw.] {10.18637/jss.v016.i04}, 16, 1

\bibitem[\protect\citeauthoryear{{Melchior} \& {Viola}}{{Melchior} \&
  {Viola}}{2012}]{2012MNRAS.424.2757M}
{Melchior} P.,  {Viola} M.,  2012, \mn@doi [\mnras]
  {10.1111/j.1365-2966.2012.21381.x}, \href
  {http://adsabs.harvard.edu/abs/2012MNRAS.424.2757M} {424, 2757}

\bibitem[\protect\citeauthoryear{{Refregier}, {Kacprzak}, {Amara}, {Bridle}  \&
  {Rowe}}{{Refregier} et~al.}{2012}]{2012MNRAS.425.1951R}
{Refregier} A.,  {Kacprzak} T.,  {Amara} A.,  {Bridle} S.,   {Rowe} B.,  2012,
  \mn@doi [\mnras] {10.1111/j.1365-2966.2012.21483.x}, \href
  {http://adsabs.harvard.edu/abs/2012MNRAS.425.1951R} {425, 1951}

\bibitem[\protect\citeauthoryear{{Simon} \& {Schneider}}{{Simon} \&
  {Schneider}}{2016}]{2016arXiv160907937S}
{Simon} P.,  {Schneider} P.,  2016, preprint, \href
  {http://adsabs.harvard.edu/abs/2016arXiv160907937S} {} (\mn@eprint {arXiv}
  {1609.07937})

\bibitem[\protect\citeauthoryear{{Viola}, {Kitching}  \& {Joachimi}}{{Viola}
  et~al.}{2014}]{2014MNRAS.439.1909V}
{Viola} M.,  {Kitching} T.~D.,   {Joachimi} B.,  2014, \mn@doi [\mnras]
  {10.1093/mnras/stu071}, \href
  {http://adsabs.harvard.edu/abs/2014MNRAS.439.1909V} {439, 1909}

\bibitem[\protect\citeauthoryear{{Voinov}}{{Voinov}}{1985}]{Voinov:1985un}
{Voinov} V.~G.,  1985, Sankhya B, 47, 354

\bibitem[\protect\citeauthoryear{{de Vaucouleurs}}{{de
  Vaucouleurs}}{1948}]{1948AnAp...11..247D}
{de Vaucouleurs} G.,  1948, Ann. Astrophys., \href
  {http://adsabs.harvard.edu/abs/1948AnAp...11..247D} {11, 247}

\makeatother
\end{thebibliography}

\appendix

\section{A ratio estimator for normal variates}
\label{sec:ratio-estimator}

Let~$x$ and~$y$ be jointly normal variates with unknown means~$\mu_x$ and~$\mu_y$, and known variances~$\sigma_x^2$ and~$\sigma_y^2$ and correlation~$\rho$.
The goal here is to find an unbiased estimate of the ratio~$r$ of their means,
\begin{equation}\label{eq:r}
	r = \mu_x/\mu_y \;.
\end{equation}
Under the additional assumption that the sign of~$\mu_y$ is known, an unbiased estimator for~$r$ can be found in two short steps.

First, the transformation of \citet{Marsaglia:1965wx,Marsaglia:2006be} is used to construct a new variate,
\begin{equation}\label{eq:w}
	w = x - c \, y \;, \quad
	c = \rho \, \sigma_x/\sigma_y \;.
\end{equation}
The constant~$c$ is chosen so that $\E[w y] = \E[w] \E[y]$.
It is clear that $w$ is normal, and that variates $w$ and $y$ are jointly normal, uncorrelated, and thus independent.
Note that $c = 0$ and $w = x$ for independent $x$ and $y$.

Secondly, \citet{Voinov:1985un} derived an unbiased estimator for the inverse mean of the normal variate~$y$, i.e.\@ a function~$\hat g(y)$ with~$\E[\hat g(y)] = 1/\mu_y$.
For the relevant case of an unknown but positive mean $\mu_y > 0$, this function is given by
\begin{equation}\label{eq:hat-g}
	\hat g(y)
	= \frac{\sqrt{\pi}}{\sqrt{2} \, \sigma_y} \, 
	    \exp\bigg(\frac{y^2}{2 \, \sigma_y^2}\bigg) \, 
	    \erfc\bigg(\frac{y}{\sqrt{2} \, \sigma_y}\bigg) \;.
\end{equation}
It is then straightforward to construct an estimator for the ratio~\eqref{eq:r},%
\begin{equation}\label{eq:hat-r}
	\hat r = c + w \, \hat g(y) \;.
\end{equation}
Since $w$ and $y$ are independent, the expectation is
\begin{equation}
	\E[\hat r]
	= c + \E[w] \E[\hat g(y)]
	= c + (\mu_x - c \, \mu_y)/\mu_y
	= \mu_x/\mu_y \;,
\end{equation}
which shows that $\hat r$ is in fact an unbiased estimator for $r$.

The variance of the ratio estimator~$\hat r$ is formally given by
\begin{equation}
	\Var[\hat r]
	= \big({\E[w]}^2 + \Var[w]\big) \Var[\hat g(y)]
	  + \Var[w] \, \E[\hat g(y)]^2 \;.
\end{equation}
As pointed out by \citet{Voinov:1985un}, the variance~$\Var[\hat g(y)]$ does not exist for function~\eqref{eq:hat-g} due to a divergence at infinity.
The confidence interval~$h$ with probability~$p$,
\begin{equation}
	\Pr\big(|\hat g(y) - 1/\mu_y| < h\big) = p \;,
\end{equation}
however, is well-defined, and the variance of~$\hat g(y)$ remains finite in applications where infinite values of $y$ are not observed.
In this case, an unbiased estimator
\begin{equation}
	\hVar[\hat g(y)]
	= \big(\hat g(y)\big)^2
	  - \big(1 - y \, \hat g(y)\big)/\sigma_y^2
\end{equation}
exists and, together with Voinov's estimator for~$1/\mu_y^2$, yields
\begin{equation}\label{eq:var-r}
	\hVar[\hat r]
	= w^2 \, \big(\hat g(y)\big)^2
	  - \big((w^2 - \sigma_x^2)/\sigma_y^2 + c^2\big) \,
	      \big(1 - y \, \hat g(y)\big)
\end{equation}
as an unbiased estimate of the variance of the estimator~$\hat r$.

When $y$ is significantly larger than its standard deviation, e.g.\@~$y/\sigma_y > 10$, the function~$\hat g(y)$ given by~\eqref{eq:hat-g} is susceptible to numerical overflow.
However, in this regime, it is also very well approximated by its series expansion,
\begin{equation}
	\hat g(y)
	\approx 1/y - \sigma_y^2/y^3 + 3 \, \sigma_y^4/y^5 \;, \quad 
	y/\sigma_y > 10 \;.
\end{equation}
For even larger values $y \gg \sigma_y$, this approaches $1/y$, as expected.

\bsp	% typesetting comment
\label{lastpage}
\end{document}